\documentclass[english]{IEEEtran}
\usepackage[T1]{fontenc}
\usepackage[latin9]{inputenc}
\usepackage{array}
\usepackage{float}
\usepackage{booktabs}
\usepackage{graphicx}

\makeatletter

\providecommand{\tabularnewline}{\\}

\usepackage{placeins}
\usepackage{balance}

\makeatother

\usepackage{babel}
\begin{document}
\title{RF Signal Classification with Synthetic Training Data and its Real-World
Performance}
\author{Stefan Scholl \\
research@panoradio-sdr.de \\
}
\maketitle
\begin{abstract}
Neural nets are a powerful method for the classification of radio
signals in the electromagnetic spectrum. These neural nets are often
trained with synthetically generated data due to the lack of diverse
and plentiful real RF data. However, it is often unclear how neural
nets trained on synthetic data perform in real-world applications.
This paper investigates the impact of different RF signal impairments
(such as phase, frequency and sample rate offsets, receiver filters,
noise and channel models) modeled in synthetic training data with
respect to the real-world performance. For that purpose, this paper
trains neural nets with various synthetic training datasets with different
signal impairments. After training, the neural nets are evaluated
against real-world RF data collected by a software defined radio receiver
in the field. This approach reveals which modeled signal impairments
should be included in carefully designed synthetic datasets. The investigated
showcase example can classify RF signals into one of 20 different
radio signal types from the shortwave bands. It achieves an accuracy
of up to 95\% in real-world operation by using carefully designed
synthetic training data only.
\end{abstract}

\section{Introduction}

This paper investigates deep learning for RF signal classification
with the application to RF signal identification. Signal identification
is the task to identify the type of an unknown wireless signal in
the electromagnetic spectrum. The ``type'' of a signal is sometimes
also called transmission mode or service. It is usually defined by
some wireless standard with which the waveform is generated (e.g.
WiFi, Bluetooth, AM radio broadcast, Stanag 4285, Morse code).

\begin{figure}
\begin{centering}
\includegraphics[scale=0.7]{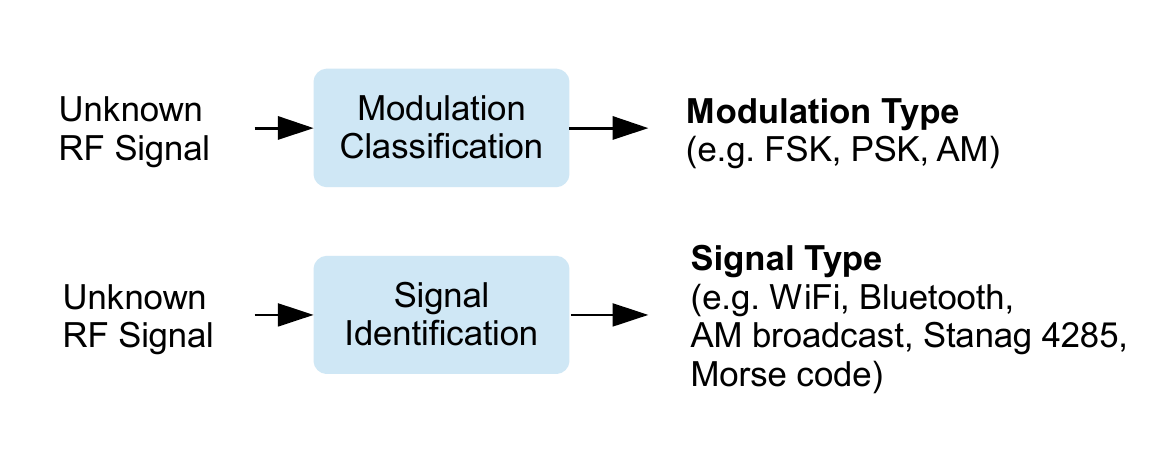}
\par\end{centering}
\centering{}\caption{\label{fig:AMC-vs-classification}Signal identification vs automatic
modulation classification (AMC): AMC outputs only the bare modulation
type, whereas signal identification determines exactly the actually
sent signal type or mode following some wireless standard.}
\end{figure}

Unlike pure automatic modulation classification (AMC), which extracts
only the modulation itself (e.g. PSK, FM, FSK), signal identification
also needs to incorporate other characteristic signal parameters such
as baud rate, shaping, frame structure or signal envelope to identify
the signal type, see Figure \ref{fig:AMC-vs-classification}. Nevertheless,
the methods used to solve AMC and signal identification tasks are
highly related.

Signal classification is used for spectrum sensing, e.g. to enable
dynamic spectrum access in cognitive radio, for spectrum monitoring
and signal intelligence applications. 

In recent years, machine learning methods like deep neural nets have
emerged as a powerful approach to classification problems in the radio
domain \cite{Wong.01.10.2020,Scholl.11.06.2019,Wu.2020,OShea.2016,TimothyJ.OShea.2017b}.
Unlike classical algorithm design, which depends on the designers
experience and his ability to recognize characteristic signal patterns,
machine learning uses large amounts of data to extract meaningful
features automatically in a training process.

The challenge of many machine learning approaches is the collection
of plenty and diverse training data, which is a requirement for creating
powerful systems. The two main types of data are real-world data and
synthetic data. Collecting RF training data from a real-world operational
scenario often requires a high effort and lacks diversity. This is
because the data obtained in a measurement campaign may be specific
to the used receiver hardware, current channel conditions and the
currently present transmitters. When real data is used, it is often
collected in the lab \cite{TimothyJ.OShea.2017b,Bitar.2017,Schmidt.2017}
and possibly lacks some real-world effects.

An alternative to real-world data is synthetically generated data.
Synthetic data consists of computer-generated waveforms, that are
distorted by a channel simulator in software. The channel simulator
adds various impairments to the pure signal waveform to model very
different reception scenarios, that possibly occur in real-world operation.
This impairments can include e.g. frequency offsets, addition of noise
or the introduction of fading. The advantage of synthetic data is
twofold:
\begin{itemize}
\item It can be generated in large amounts
\item Diverse signal impairments can be modeled, that lead to robust neural
nets, that generalize well.
\end{itemize}
The main drawback of synthetic training is that it is often unclear
how the signals need to be distorted in order to obtain a good classification
accuracy in practical operation: Which channel models and effects
are important, which are not? How to parameterize the channel simulator
in order to obtain realistic, but also diverse data, that ideally
models all receiver situations encountered in the real-world application?

\section{Contribution \& Related Work}

Some papers have investigated the effects of a few isolated channel
impairments in synthetic training data, such as frequency offsets
\cite{Hauser.2017} or the proper choice of SNR values \cite{SharanRamjee.2019}.
However, this covers only a small fraction of the relevant signal
impairments. Furthermore, they do not investigate the effects with
real-world data. Other works \cite{WilliamH.ClarkIV.2020,TimothyJ.OShea.2017b}
use real-world data for validation, but do not investigate the impact
of specific signal impairments in the synthetic dataset.

This paper investigates the influence of synthetic training data on
the real-world performance of a signal classifier, thus linking synthetic
training with real-world classification performance. For that purpose,
several different datasets with varying amount of signal distortion
are generated. For each dataset a neural network is trained and evaluated
to monitor its performance in a real-world application. Comparing
the real-world classification performance reveals the influence of
different simulated signal distortions in the training data. In addition,
the results provide insight in the generalization ability of the network.

\section{Training Dataset Generation}

\subsection{Signal Classes}

All training datasets contain 20 different signal types, that are
shown in Table \ref{tab:training_dataset_modes}. This set includes
a large number of digitally modulated signals, such as radioteletype,
Navtex, PSK modes, multiple FSK and multi-carrier modes. In addition,
different analog modulated signals, such as AM broadcasting, single-sideband
(SSB) voice and HF fax are included. These signals are commonly used
by commercial, amateur and governmental operators across the shortwave
band between 3-30 MHz.

\subsection{Dataset}

A complete dataset contains 120,000 synthetically generated signals
for training and another 30,000 for training validation. Each signal
consists of 2048 complex IQ samples with a sampling frequency of 6
kHz. This results in a signal duration of approximately 340 ms.

\begin{table}
\begin{centering}
\begin{tabular}{lll}
\toprule 
Mode Name & Modulation & Baud Rate\tabularnewline
\midrule
PSK31 & PSK & 31\tabularnewline
PSK63 & PSK & 63\tabularnewline
RTTY 45/170 & FSK, 170 Hz shift & 45\tabularnewline
RTTY 50/450 & FSK, 450 Hz shift & 50\tabularnewline
RTTY 75/170 & FSK, 170 Hz shift & 75\tabularnewline
Navtex / Sitor-B & FSK, 170 Hz shift & 100\tabularnewline
Olivia 4/500 & 4-MFSK & 125\tabularnewline
Olivia 8/250 & 8-MFSK & 31\tabularnewline
Olivia 16/500 & 16-MFSK & 31\tabularnewline
Olivia 32/1000 & 32-MFSK & 31\tabularnewline
Contestia 16/250 & 16-MFSK & 16\tabularnewline
MFSK-16 & 16-MFSK & 16\tabularnewline
MFSK-32 & 16-MFSK & 31\tabularnewline
MFSK-64 & 16-MFSK & 63\tabularnewline
MT63 / 500 & Multi-Carrier & 5\tabularnewline
USB (voice) & Single-Sideband (upper) & analog\tabularnewline
LSB (voice)  & Single-Sideband (lower) & analog\tabularnewline
AM broadcast & AM & analog\tabularnewline
Morse Code & OOK & variable\tabularnewline
HF / Radio Fax & Facsimile & analog\tabularnewline
\bottomrule
\end{tabular}
\par\end{centering}
\medskip{}

\caption{\label{tab:training_dataset_modes}Signal classes in the datasets}
\end{table}

\subsection{Modeled Effects}

\begin{figure}
\begin{centering}
\includegraphics[scale=0.6]{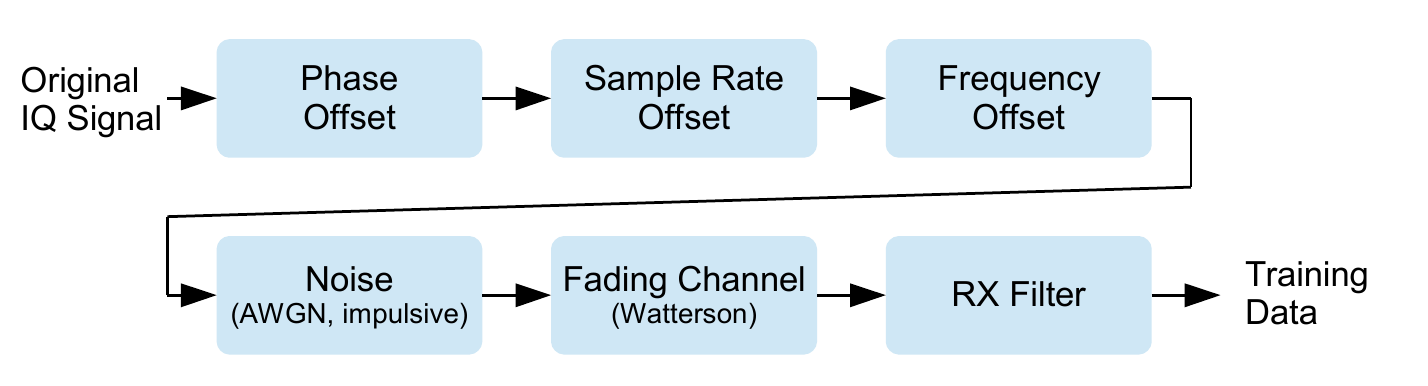}
\par\end{centering}
\caption{\label{fig:Synthetic-Dataset-Generation}Synthetic dataset generation}
\end{figure}

An overview of all modeled signal impairments is depicted in Figure
\ref{fig:Synthetic-Dataset-Generation}. In the following the impairments
are explained in detail.

\subsubsection{Phase and Frequency Offset}

In general, transmitter and receiver phase and frequency are not synchronized
when the signal type is unknown (blind reception). Therefore, a random
constant phase shift is added to each signal in the dataset. In addition,
a frequency offset is introduced, that is randomly chosen between
-250 and +250 Hz to model receiver tuning mismatch.

\subsubsection{Sample Rate Offset}

The sample clock of transmitter and receiver electronics are usually
not exactly synchronized. Therefore, a small difference in sample
frequencies may be present, that virtually stretches or compresses
the signal by a small amount. The generated data uses randomly selected
sample frequency offsets of 0, +/-0.5 and +/-1\%.

\subsubsection{Additive Noise}

Receiver noise is typically modeled as additive white Gaussian noise
(AWGN). However, also other noise types may be present, such as impulsive
noise. Impulsive noise often originates from lightnings and other
interference in shortwave channels. Thus impulsive noise may sometimes
model the noise more accurately. In this paper, impulsive noise is
generated by exponentiating the magnitude of ordinary AWGN noise,
i.e.

\[
n_{impulsive}(t)=sign\left(n_{AWGN}(t)\right)\cdot|n_{AWGN}(t)|^{x}
\]

\begin{figure}
\begin{centering}
\includegraphics[scale=0.43]{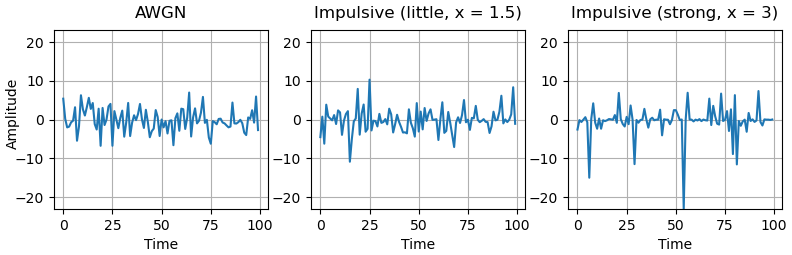}
\par\end{centering}
\caption{\label{fig:AWGN-vs-impulsive}AWGN vs impulsive noise (same noise
power)}
\end{figure}

During data generation, the noise type is selected randomly among
AWGN and impulsive noise with exponents $x=1.5,2$ or $3$ and added
to the signal (see Figure \ref{fig:AWGN-vs-impulsive}).

Proper selection of the noise power ensures that the desired SNR value
is achieved. Throughout this paper SNR is referred to the complete
Nyquist bandwidth.

\subsubsection{\label{subsec:Fading-Channels-for}Fading Channels for Shortwave}

\begin{figure}
\begin{centering}
\includegraphics[scale=0.55]{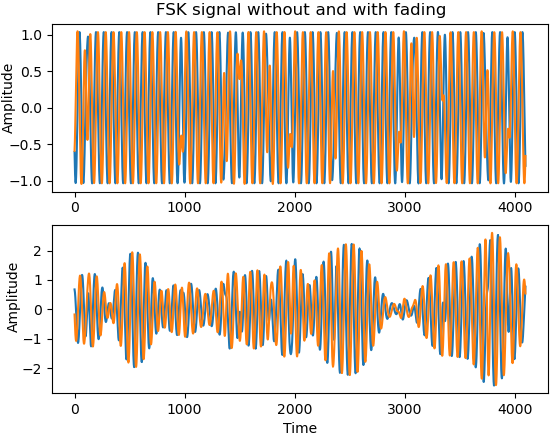}
\par\end{centering}
\caption{\label{fig:fading-vs-no-fading}The effect of fading on the amplitude
of a RTTY 45/170 IQ signal}
\end{figure}

Time-varying multi-path propagation is present in many communication
channels and manifests in fading effects. Fading results in fluctuations
of the amplitude of the received signal over time (Figure \ref{fig:fading-vs-no-fading})
and frequency (Appendix, Figure \ref{fig:fading-examples}). At shortwave
frequencies, communication mainly occurs through reflections of the
electromagnetic waves at the ionosphere. This ionospheric propagation
introduces fading and varying Doppler shifts.

\begin{figure}
\begin{centering}
\includegraphics[scale=0.75]{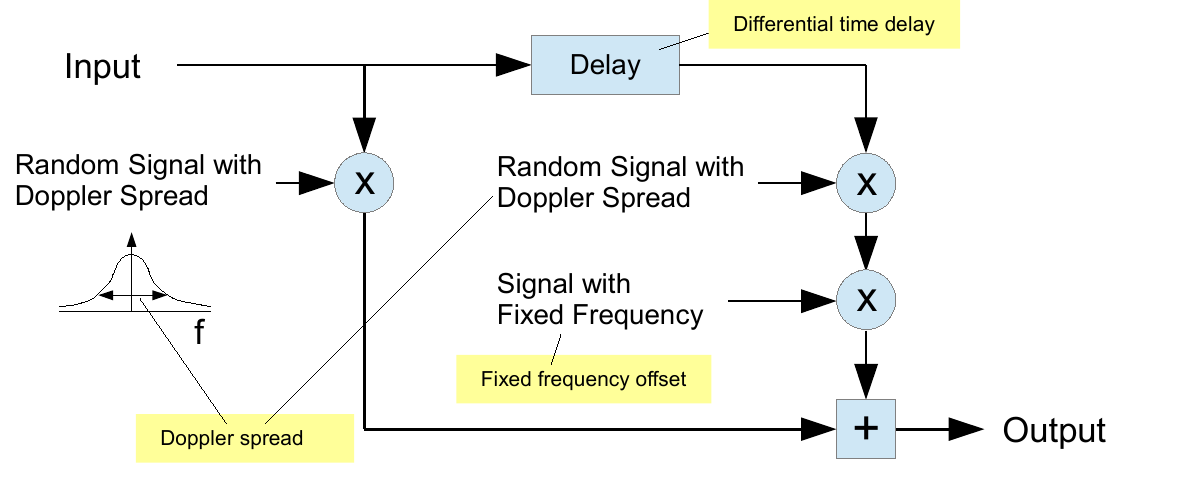}
\par\end{centering}
\caption{\label{fig:Watterson-Fading-Model}Watterson fading model as used
by the HF channel standards CCIR 520 and ITU 1487}

\end{figure}

The Watterson model \cite{C.Watterson.1970} is a widely used model
for ionospheric propagation and is depicted in Figure \ref{fig:Watterson-Fading-Model}.
It consists of a two-tap delay line, that models two different propagation
paths. Each path is multiplied by a independent random noise-like
signal, that is frequency filtered such that is introduces random
Doppler shifts. In addition, a fixed frequency offset can be introduced
in one of the two paths. A concrete realization of the Watterson model
is defined by three parameters: differential delay (in ms), Doppler
spread (in Hz) and fixed frequency offset (in Hz) (see Figure \ref{fig:Watterson-Fading-Model}).
Different values of these three parameters correspond to different
propagation behaviors.

Two standards have been developed, that follow this definition of
the Watterson model: CCIR 520 \cite{CCIR.1992} and ITU 1487 \cite{ITUR.052000}.
CCIR 520 defines several parameter sets for channels described by
their quality (``flat'', ``good'', ``moderate'', ``poor'',
``flutter'' and ``doppler''). ITU 1487 extends the number of parameter
sets and introduces channel models for locations at low, medium and
high latitudes. Moreover, the parameters of the Watterson model can
be set to values that deviate from the standards CCIR and ITU in order
to generate more extreme models. This may be useful for training data
generation, because additional extreme channel data can improve the
generalization capability of the trained network. In summary, this
paper investigates three sets of channel models:
\begin{itemize}
\item CCIR 520
\item ITU 1487
\item Extended channel set (CCIR, ITU and extreme models)
\end{itemize}
Further details on the three sets of fading channel models can be
found in the Appendix.

Each signal in the dataset is distorted by a channel model, that is
randomly selected from the set of channel models (including the option
to use no channel model). 

\subsubsection{Receiver Filter (RX Filter)}

\begin{figure}
\begin{centering}
\includegraphics[scale=0.55]{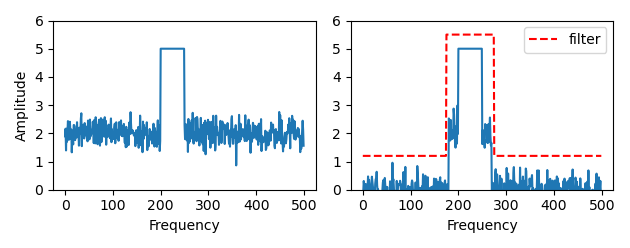}
\par\end{centering}
\caption{\label{fig:rx_filter_fig}Signal before (left) and after applying
a bandpass filter in the receiver (right)}

\end{figure}

Signals encountered in practical operation often have varying bandwidths.
In the receiver, the operator or an automatic algorithm usually applies
a filter, that reduces noise outside the signal bandwidth prior to
classification. Therefore a classifier needs to deal with the fact
that the noise may not equally be distributed across the complete
Nyquist bandwidth as depicted in Figure \ref{fig:rx_filter_fig}.
This effect is modeled in the dataset by applying a bandpass filter.
In practice it is not always possible to set a receiver filter exactly
to the signal's bandwidth, e.g. because a signal's bandwidth is not
properly defined or the SNR is very low. Therefore the filter's bandwidth
in the training data varies between the signal bandwidth and the full
Nyquist bandwidth in order to train the network on different receiver
filter bandwidths.

\section{\label{sec:Real-World-Validation-Data}Real-World Validation Data}

Most scientific works that use synthetic data for training, also use
synthetic data of the same distribution for validation (e.g. in \cite{Wu.2020,OShea.2016,SharanRamjee.2019}).
Therefore such ordinary synthetic validation data can only validate
the training process and not the final performance in the real application.
This paper uses a non-synthetic dataset for validation. It consists
of real-world signals, that have been captured ``in the field''.
This real-world validation data allows to assess the network's final
performance under real-world conditions.

The real-world signals have been obtained using the Twente University
WebSDR \cite{PieterTjerkdeBoer.UniversityofTwente} receiver. The
recorded signals are signals of opportunity coincidentally present
at reception time in the spectrum. The recorded signals have a high
diversity: The signals have been recorded at numerous different SNRs,
different times of day and season and different frequencies. This
ensures the occurrence of largely different channel conditions (Since
channel conditions in the shortwave band are highly dependent on day,
season and sun activity). Furthermore, most of the signals were captured
from different transmitting stations. This mitigates biasing effects
that may originate in emitter specific properties, such as location,
transmitter electronics and signal generation.

The receiver filter is roughly set to the signal bandwidth.

All recordings are cut into a collection of data signals with each
having a length of 2048 IQ samples (same as training data format).
Apart from that, no further change of the signal is made. Thus all
signal impairments remain ``as they are''.

Table \ref{tab:real-world-validation-data} presents an overview of
the recorded data. The total recording time of the real-world signals
is in the order of several hours. Since the recorded signal types
have different amounts of data, it is important to balance the classes
properly in order to get meaningful results.

\begin{table}
\begin{centering}
\begin{tabular}{>{\raggedright}p{2cm}ll>{\raggedright}p{2.8cm}}
\toprule 
Mode Name & \# Data & SNR Range & Frequencies / Bands\tabularnewline
\midrule
PSK31 & 3,700 & -10 to +25 dB & 40 m, 20 m\tabularnewline
PSK63 & 5,500 & -10 to +25 dB & 80 m, 40 m, 20 m\tabularnewline
RTTY 45/170 & 5,200 & -10 to +40 dB & 40 m, 20 m\tabularnewline
RTTY 50/450 & 11,300 & -10 to +35 dB & 4.5 to 11.0 MHz\tabularnewline
RTTY 75/170 & 3,600 & -10 to +35 dB & 80 m, 40 m, 20 m\tabularnewline
Navtex / Sitor-B & 9,000 & -10 to +20 dB & 0.5, 4.2, 8.4, 12.6 MHz\tabularnewline
Olivia 4/500 & 1,300 & +5 to +15 dB & 80 m\tabularnewline
Olivia 8/250 & 7,600 & -10 to +25 dB & 80 m to 20 m\tabularnewline
Olivia 16/500 & 4,000 & -10 to +20 dB & 80 m, 40 m, 20 m\tabularnewline
Olivia 32/1000 & 4,700 & -5 to +35 dB & 20 m\tabularnewline
Contestia 16/250 & 1,700 & -10 to +10 dB & 40 m, 20 m\tabularnewline
MFSK-16 & 3,500 & -10 to +15 dB & 80 m, 40 m, 20 m\tabularnewline
MFSK-32 & 4,000 & -10 to +15 dB & 19 m, 7.8 MHz\tabularnewline
MFSK-64 & 5,400 & -10 to +15 dB & 19 m, 7.8 MHz\tabularnewline
MT63 / 500 & 1,300 & -10 to +15 dB & 40 m\tabularnewline
USB (voice) & 12,700 & -10 to +30 dB & 5 to 28 MHz\tabularnewline
LSB (voice) & 13,100 & -10 to +25 dB & 3 to 10 MHz\tabularnewline
AM broadcast & 10,900 & -10 to +45 dB & 75 m to 16 m\tabularnewline
Morse Code & 6,300 & -10 to +20 dB & 80 m to 10 m\tabularnewline
HF / Radio Fax & 20,100 & -10 to +35 dB & 3.9 to 13.8 MHz\tabularnewline
\bottomrule
\end{tabular}
\par\end{centering}
\medskip{}

\caption{\label{tab:real-world-validation-data}Recorded real-world validation
data for final evaluation: data from various transmitters has been
recorded at different times of day, season, frequency and location}
\end{table}

\section{Training}

For the evaluation of the synthetic dataset generation, eleven separate
datasets with different properties have been generated. Table \ref{tab:training_datasets_4result}
shows an overview of the all datasets. From the top down, each dataset
adds an additional signal impairment. The first dataset contains no
impairments, whereas the last dataset contains all considered impairments.
Note, that the dataset names indicate the signal impairment that has
been specifically added to this dataset (see Table \ref{tab:training_datasets_4result}).

\begin{table*}
\begin{centering}
\begin{tabular}{lllll>{\raggedright}p{1cm}l}
\toprule 
Dataset Name & Frequency Offset & Phase Offset & fs Offset & Noise & RX Filter & Fading Channel\tabularnewline
\midrule
No Augmentation & - & - & - & - & - & -\tabularnewline
Frequency Offset & \textbf{+/- 250 Hz} & - & - & - & - & -\tabularnewline
Phase Offset & +/- 250 Hz & \textbf{random} & - & - & - & -\tabularnewline
fs Offset & +/- 250 Hz & random & \textbf{yes} & - & - & -\tabularnewline
\midrule
AWGN, high SNR & +/- 250 Hz & random & yes & \textbf{AWGN: +5 to +25 dB SNR} & - & -\tabularnewline
AWGN, full SNR & +/- 250 Hz & random & yes & AWGN: \textbf{-15 to +25 dB SNR} & - & -\tabularnewline
Impulsive Noise & +/- 250 Hz & random & yes & AWGN or \textbf{Impulsive}: -15 to +25 dB SNR & - & -\tabularnewline
\midrule
RX Filter & +/- 250 Hz & random & yes & AWGN or Impulsive: -15 to +25 dB SNR & \textbf{yes} & -\tabularnewline
\midrule
CCIR Fading & +/- 250 Hz & random & yes & AWGN or Impulsive: -15 to +25 dB SNR & yes & \textbf{CCIR 520}\tabularnewline
ITU Fading & +/- 250 Hz & random & yes & AWGN or Impulsive: -15 to +25 dB SNR & yes & \textbf{ITU 1487}\tabularnewline
Extended Fading & +/- 250 Hz & random & yes & AWGN or Impulsive: -15 to +25 dB SNR & yes & \textbf{Extended}\tabularnewline
\bottomrule
\end{tabular}
\par\end{centering}
\medskip{}

\caption{\label{tab:training_datasets_4result}Analyzed training datasets}
\end{table*}

\begin{figure}
\begin{centering}
\includegraphics[scale=0.5]{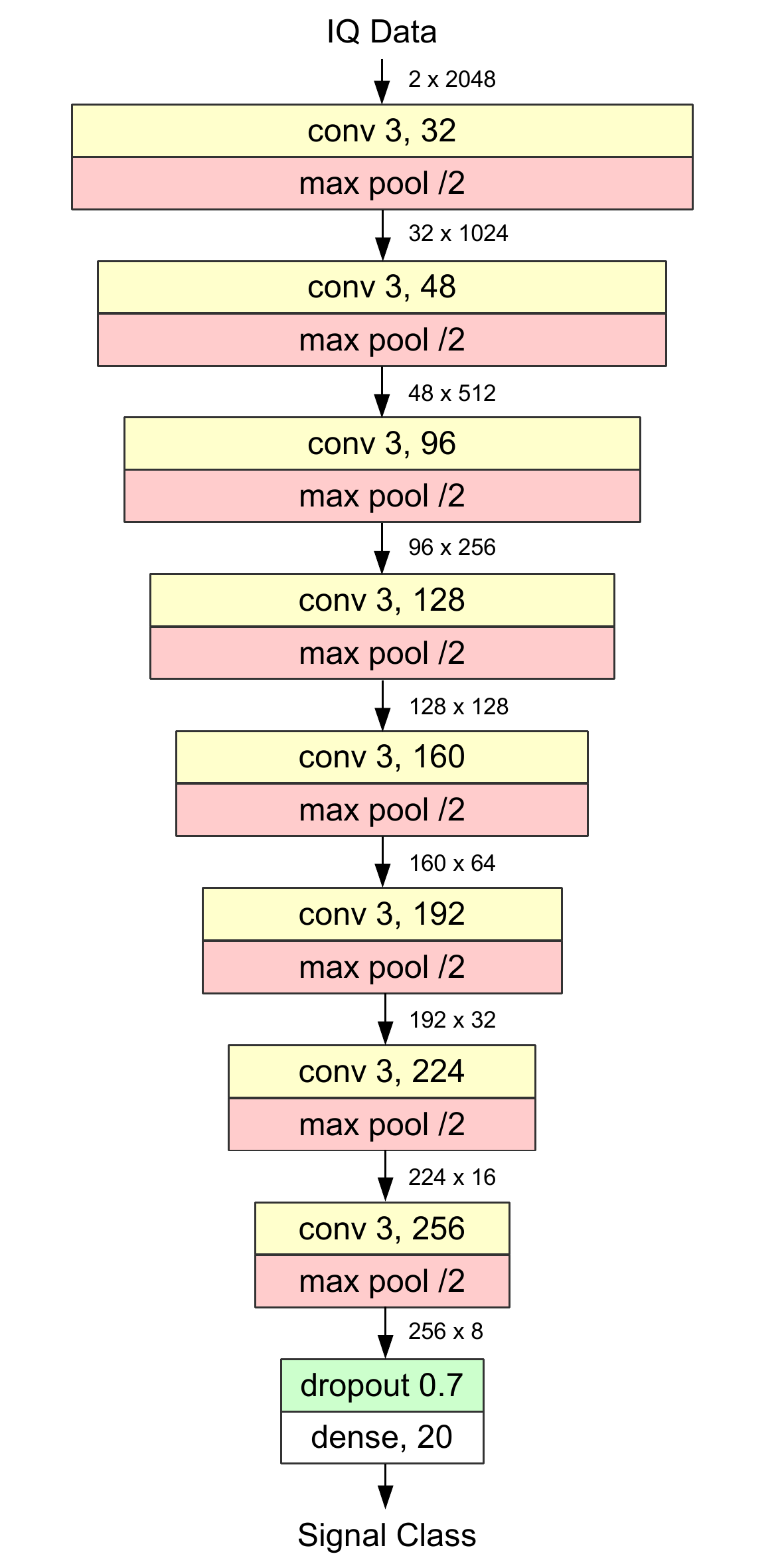}
\par\end{centering}
\caption{\label{fig:CNN}9-layer CNN used for training}

\end{figure}

The synthetic datasets are used to train the neural network shown
in Figure \ref{fig:CNN}. The network is a convolutional neural network
(CNN) with 9 layers and approximately 550,000 parameters. The complex
IQ input data is fed into the network with real and imaginary part
separated into two input channels. The CNN is trained on 120,000 signals
using an Adam optimizer and a batch size of 128 for 20 epochs.

After the training is completed, the resulting trained net is finally
validated once with the real-world validation dataset to measure its
performance in a real-world scenario.

To mitigate statistical effects of the training process, five networks
have been trained independently for each dataset and the average accuracy
is considered as final result.

\section{Results}

\subsection{Overall}

\begin{figure}
\begin{centering}
\includegraphics[scale=0.6]{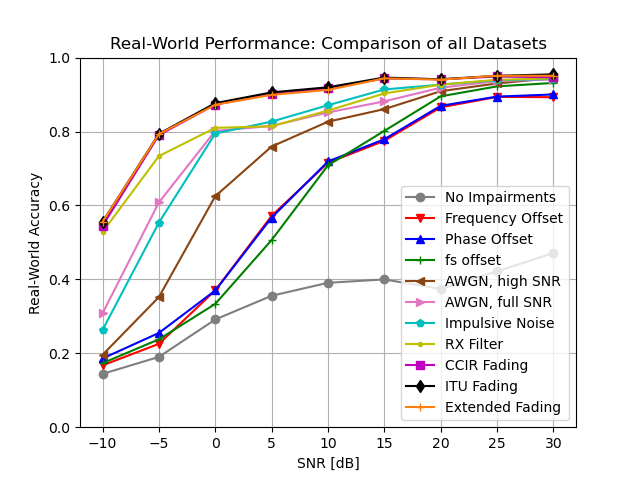}
\par\end{centering}
\caption{\label{fig:main_results}Comparison of the real-world performance
for all eleven training datasets}
\end{figure}

Figure \ref{fig:main_results} presents the main results of the paper.
It shows the accuracy of the neural networks, each trained by one
of the eleven different datasets measured against the real-world dataset.
The most sophisticated training datasets achieve an accuracy of 95
\% for high SNR values. This demonstrates, that training with carefully
designed synthetic data generalizes well to real-world data. Next,
the impact of the different channel impairments is investigated in
greater detail.

\subsection{Impact of Offsets in Frequency, Phase and Sample Rate}

\begin{figure}
\begin{centering}
\includegraphics[scale=0.6]{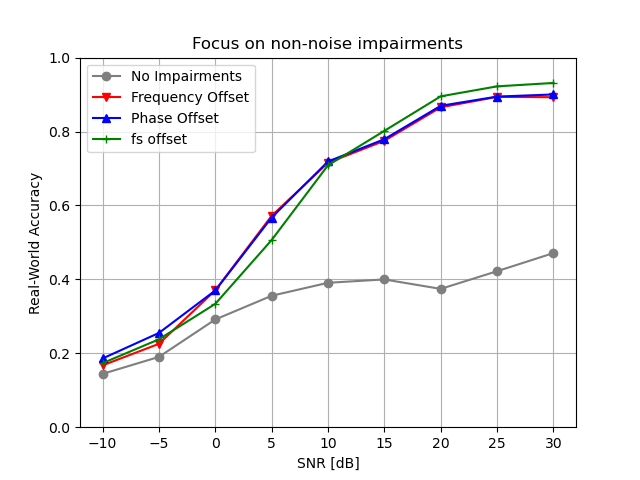}
\par\end{centering}
\caption{\label{fig:Focus-on-Non-Noise}Focus on non-noise effects: frequency,
phase and sample rate offset }
\end{figure}

Figure \ref{fig:Focus-on-Non-Noise} focuses on the impact of frequency,
phase and sample rate offset. As a first interesting observation,
training with clean signals without any impairments works remarkably
well and accuracies above 40\% can be obtained. Introducing frequency
offsets to the training data largely improves accuracy. This is because
adding a frequency offset can heavily change the signal shape in the
time domain. The introduction of phase offsets has very little impact,
showing that the network does not respond to absolute phase. The application
of sample rate offsets provides a minor improvement for high SNR at
the expense of slightly lower accuracy at smaller SNR.

\subsection{Impact of AWGN and Impulsive Noise}

\begin{figure}
\begin{centering}
\includegraphics[scale=0.6]{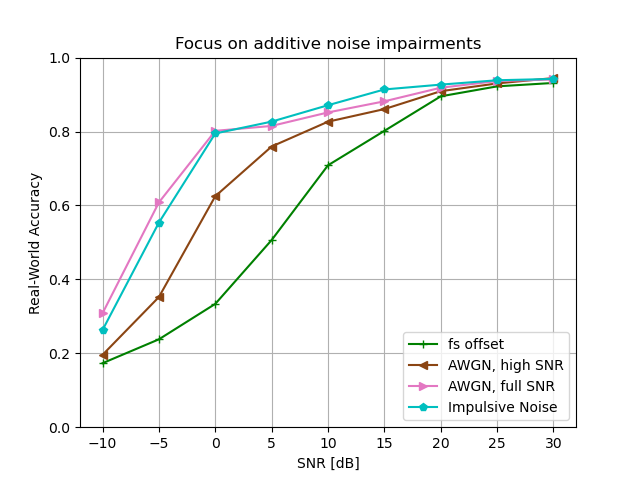}
\par\end{centering}
\caption{\label{fig:Focus-on-Noise}Focus on additive noise effects}
\end{figure}

The question how noise impacts the final accuracy can be answered
with Figure \ref{fig:Focus-on-Noise}. The baseline is the ``fs offset''
dataset containing the above mentioned non-noise effects, but no noise
impairments. The training dataset ``AWGN, high SNR'' additionally
introduces AWGN noise to the data with SNR values between +5 to +25
dB (see Table \ref{tab:training_datasets_4result}). As expected this
improves the accuracy above 5 dB SNR. However, it also improves accuracy
below 5 dB although low SNR data is not yet included in the training
data. This is strong evidence that the network is able to generalize
towards lower SNR.

The training dataset ``AWGN, full SNR'' introduces AWGN noise with
a wider SNR range from -15 dB to +25 dB, i.e. it also includes low
SNR values. The inclusion of lower SNR values improves the accuracy
in the low SNR region below 5 dB as expected. However, also a small
increase at SNRs above 5 dB can be observed, which again is evidence
that the network improved its generalization capabilities.

The inclusion of impulsive noise showed no major advantage. The accuracy
slightly increases for higher SNR and decreases for lower SNR. Although
impulsive noise does not provide a clear improvement, it may nevertheless
be useful to improve network generalization and robustness.

\subsection{Impact of Fading and RX Filtering}

\begin{figure}
\begin{centering}
\includegraphics[scale=0.6]{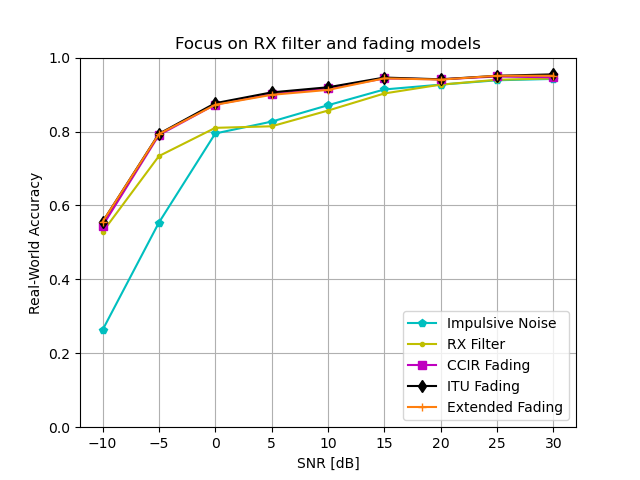}
\par\end{centering}
\caption{\label{fig:Focus-on-RX_fading}Focus on RX filtering and fading models:
CCIR 520, ITU 1487 and ``Extended''}
\end{figure}

Figure \ref{fig:Focus-on-RX_fading} shows the results for training
data, that employ receiver (RX) filtering and different fading channels.
The inclusion of the filter largely improves the accuracy in the low
SNR region. Note, that the reason for this improvement is not simply
the reduction of noise in the signal by the filter, because this filtering
has been added specifically to the training data. The real-world validation
data always includes a more or less tight filter (independent of the
training dataset), because RF receivers typically use filters to remove
out-of-band noise and interference. Since the impact of the RX filter
on the time domain signal is more present for low SNR, the improvement
becomes apparent in the low SNR region.

The introduction of fading channels shows a clear increase in accuracy.
It shows the importance of applying fading channel models, that model
amplitude fluctuations and frequency selective attenuation. Interestingly,
all three applied fading channel models (CCIR 520, ITU 1487 and the
extended custom set) provide very similar accuracy.

\section{Conclusion}

This paper demonstrated how synthetically generated training data
can be used to create radio signal classifiers with high accuracy
under real-world operation. It has been investigated how different
signal impairments influence the real-world accuracy and the network's
ability to generalize. The results show the importance of training
with frequency offsets, a wide range of SNR values and modeling an
RX filter and fading channels. The choice of the fading channel was
uncritical with regard to the exact channel parameters. Minor to no
improvements have been observed for phase offset, sample rate offset
and impulsive noise. In the example scenario of 20 RF signal classes,
a CNN, trained only by synthetic data, has obtained an accuracy of
up to 95 \% on real-world signals. 

\bibliographystyle{ieeetr}
\bibliography{literatur}

\newpage{}

\balance

\appendix

\section{Watterson Channel Models}

The appendix provides detailed information on the fading channel models
used for training data generation. The baseline model is the Watterson
model as described in the standards CCIR 520 and ITU 1487 \cite{C.Watterson.1970,ITUR.052000,CCIR.1992}.
An concise description of the Watterson model has been provided in
Section \ref{subsec:Fading-Channels-for}.

Tables \ref{tab:Channel-Models-in-ccir520}, \ref{tab:Channel-Models-in-itu}
and \ref{tab:Custom-Extreme-Channel} provide the exact parametrization
of the three investigated channel sets: CCIR 520, ITU 1487 and Extended.
Note, that the flat fading channels of CCIR 520 consider only one
path. Also note, that some channels from CCIR 520 and ITU 1487 overlap,
i.e. they have the same parametrization (e.g. CCIR 520 ``good``
and ITU ``mid quiet''). For few channels CCIR and ITU do not provide
exact parameters but a range of parameters. In these cases either
a meaningful value has been selected or two channel models have been
considered (e.g. with CCIR 520 ``Flat 1'' and ``Flat 2``).

\begin{figure}[H]
\begin{centering}
\includegraphics[scale=0.4]{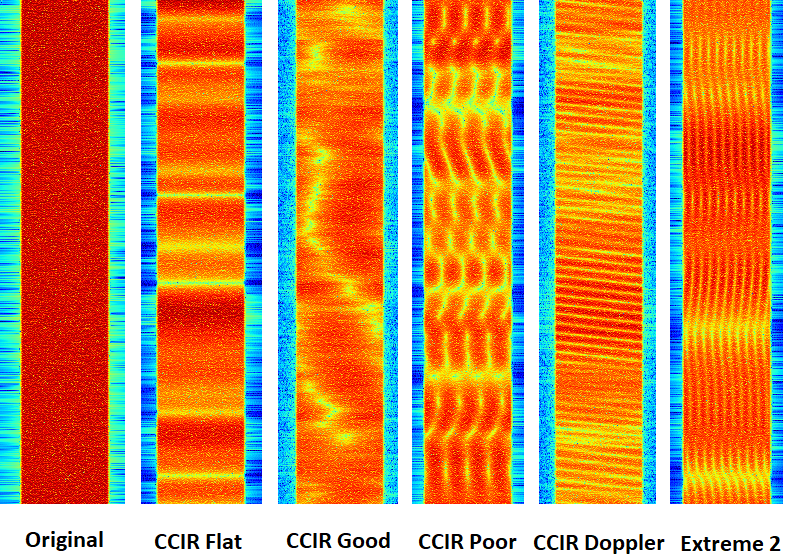}
\par\end{centering}
\caption{\label{fig:fading-examples}Example of a multi-carrier signal, that
has been modified by some of the considered Watterson channel models
(spectrogram representation, time at vertical axis)}
\end{figure}

Figure \ref{fig:fading-examples} shows the influence of some fading
models on a multi-carrier signal. It visualizes the fluctuations of
amplitude over time and frequency.

\begin{table}[H]
\begin{centering}
\begin{tabular}{l>{\raggedright}p{1cm}>{\raggedright}p{1.5cm}>{\raggedright}p{1cm}>{\raggedright}p{1cm}}
\toprule 
Channel Name & Paths / Taps & Differential Time Delay & Frequency Spread & Frequency Offset\tabularnewline
\midrule
Flat 1 & 1 & \textbf{-} & 0.2 Hz & -\tabularnewline
Flat 2 & 1 & - & 1 Hz & -\tabularnewline
Good & 2 & 0.5 ms & 0.1 Hz & -\tabularnewline
Moderate & 2 & 1 ms & 0.5 Hz & -\tabularnewline
Poor & 2 & 2 ms & 1 Hz & -\tabularnewline
Flutter & 2 & 0.5 ms & 10 Hz & -\tabularnewline
Doppler & 2 & 0.5 ms & 0.2 Hz & 5 Hz\tabularnewline
\bottomrule
\end{tabular}
\par\end{centering}
\medskip{}

\caption{\label{tab:Channel-Models-in-ccir520}Channel models defined by standard
CCIR 520}

\end{table}

\begin{table}[H]
\begin{centering}
\begin{tabular}{l>{\raggedright}p{1cm}>{\raggedright}p{1.5cm}>{\raggedright}p{1cm}>{\raggedright}p{1cm}}
\toprule 
Channel Name & Paths / Taps & Differential Time Delay & Frequency Spread & Frequency Offset\tabularnewline
\midrule
Low - Quiet & 2 & 0.5 ms & 0.5 Hz & -\tabularnewline
Low - Moderate & 2 & 2 ms & 1.5 Hz & -\tabularnewline
Low - Disturbed & 2 & 6 ms & 10 Hz & -\tabularnewline
Mid - Quiet & 2 & 0.5 ms & 0.1 Hz & -\tabularnewline
Mid - Moderate & 2 & 1 ms & 0.5 Hz & -\tabularnewline
Mid - Disturbed & 2 & 2 ms  & 1 Hz & -\tabularnewline
Mid - NVIS & 2 & 7 ms & 1 Hz & -\tabularnewline
High - Quiet & 2 & 1 ms & 0.5 Hz & -\tabularnewline
High - Moderate & 2 & 3 ms & 10 Hz & -\tabularnewline
High - Disturbed & 2 & 7 ms & 30 Hz & -\tabularnewline
\bottomrule
\end{tabular}
\par\end{centering}
\medskip{}

\caption{\label{tab:Channel-Models-in-itu}Channel models defined by standard
ITU 1487}
\end{table}

\begin{table}[H]
\begin{centering}
\begin{tabular}{>{\raggedright}p{2cm}>{\raggedright}p{0.7cm}>{\raggedright}p{0.4cm}>{\raggedright}p{1.35cm}>{\raggedright}p{0.9cm}>{\raggedright}p{1cm}}
\toprule 
Channel Name & Standard & Paths/ Taps & Differential Time Delay & Frequency Spread & Frequency Offset\tabularnewline
\midrule
Flat 1 & CCIR & 1 & \textbf{-} & 0.2 Hz & -\tabularnewline
Flat 2 & CCIR & 1 & - & 1 Hz & -\tabularnewline
Good & CCIR & 2 & 0.5 ms & 0.1 Hz & -\tabularnewline
Moderate & CCIR & 2 & 1 ms & 0.5 Hz & -\tabularnewline
Poor & CCIR & 2 & 2 ms & 1 Hz & -\tabularnewline
Flutter & CCIR & 2 & 0.5 ms & 10 Hz & -\tabularnewline
Doppler & CCIR & 2 & 0.5 ms & 0.2 Hz & 5 Hz\tabularnewline
\midrule
Low - Disturbed & ITU & 2 & 6 ms & 10 Hz & -\tabularnewline
Mid - NVIS & ITU & 2 & 7 ms & 1 Hz & -\tabularnewline
High - Moderate & ITU & 2 & 3 ms & 10 Hz & -\tabularnewline
High - Disturbed & ITU & 2 & 7 ms & 30 Hz & -\tabularnewline
\midrule
Poor Doppler & - & 2 & 2 ms & 1 Hz & 10 Hz\tabularnewline
High - Moderate Doppler & - & 2 & 3 ms & 10 Hz & 8 Hz\tabularnewline
Extreme 1 & - & 2 & 1 ms & 40 Hz & -\tabularnewline
Extreme 2 & - & 2 & 5 ms & 0.5 Hz & -\tabularnewline
\bottomrule
\end{tabular}
\par\end{centering}
\medskip{}

\caption{\label{tab:Custom-Extreme-Channel}Extended channel model set: combination
of CCIR, ITU and custom-designed extreme case channels}
\end{table}

\end{document}